# Photoinduced tunability of the Reststrahlen band in 4H-SiC

Bryan T. Spann,[1,2,†] Ryan Compton,[1,2,†] Daniel Ratchford,[1] James P. Long,[1] Adam D. Dunkelberger,[1,2] Paul B. Klein,[3] Alexander J. Giles,[1,2] Joshua D. Caldwell,[1,*] and Jeffrey C. Owrutsky[1,*]

[1]U.S. Naval Research Laboratory, Washington, D.C., 20375, USA

[2]National Research Council Postdoctoral Research Associate

[3]Sotera Defense Solutions Inc., Herndon, Virginia, 20171, USA

**ABSTRACT**

Materials with a negative dielectric permittivity (e.g. metals) display high reflectance and can be shaped into nanoscale optical-resonators exhibiting extreme mode confinement, a central theme of nanophotonics. However, the ability to *actively tune* these effects remains elusive. By photoexciting free carriers in 4H-SiC, we induce dramatic changes in reflectance near the "Reststrahlen band" where the permittivity is negative due to charge oscillations of the polar optical phonons in the mid-infrared. We infer carrier-induced changes in the permittivity required for useful tunability (~ 40 cm$^{-1}$) in nanoscale resonators, providing a direct avenue towards the realization of actively tunable nanophotonic devices in the mid-infrared to terahertz spectral range.

**TEXT**

The topic of sub-diffractional light confinement has been at the forefront of modern photonics research for more than a decade, giving rise to the field of nanophotonics. Rapid advancements in this area have prompted considerable growth in the fields of nanoscale imaging, metamaterials, and chemical/biological sensing among others. For example, the exploitation of surface plasmon polariton (SPP) resonances in metals has provided a means to achieve sub-diffractional optical confinement, leading to strong localized fields and enhanced light-matter interactions from the ultraviolet (UV) to near-infrared (NIR).[1] Alternatively, sub-diffractional confinement of light can be produced at mid-IR to THz frequencies using polar dielectric materials such as SiC,[2-6] hBN,[7-9] and GaAs[10] via the excitation of surface phonon polaritons (SPhPs).[11] Such polar dielectrics have attracted significant interest because their low optical losses yield comparatively high resonant quality factors [Q = $\omega_0/\Delta\omega$, where $\omega_0$ and $\Delta\omega$ are the resonant frequency and respective full width at half maximum (FWHM)] for nanoparticle optical-resonators. For example, theoretical Q's for a nanosphere exceed 900 in SiC compared to ~ 40 for Ag, the typical metal of choice for low-loss plasmonics.[12] In addition, as we show here, the underlying dielectric permittivity of a polar semiconductor can be dynamically and significantly altered by introducing free carriers, thus providing the promise for realizing active control of the optical and electronic properties of SPhP devices.

SPhPs arise from coherent charge oscillations supported by optical phonons of a polar lattice and can be stimulated when the real part of the material dielectric function becomes negative.[3-4, 12-13] As with metals, this negative permittivity results in high reflectivity. For polar dielectrics, this occurs in the spectral region between the longitudinal (LO) and transverse optical (TO) phonon frequencies and is referred to as the 'Reststrahlen' band, which depending on material can be found from the mid-infrared to terahertz (~6-350 µm) spectral range. For example in SiC, the LO and TO phonon frequencies occur at about ~970 cm$^{-1}$ and ~797 cm$^{-1}$, respectively, resulting in high reflectivity from ~ 10.3 µm to 12.5 µm. Analogous to SPPs, SPhP fields are evanescent in character (confined to sub-diffraction length scales near the surface) and can support both localized and propagating surface modes. Because the phenomenon is derived from optical phonons whose 1-10 picosecond scattering lifetimes are several orders of magnitude longer than free carrier scattering, the optical losses are drastically lower for SPhPs, resulting in narrow resonance linewidths and high Q factors for sub-diffractional polar dielectric nanostructures.[2-3, 5-6, 8, 12] This has recently been demonstrated with experimental values extending upwards to ~300 in SiC[14] and hBN[15] nanostructures.

One distinct advantage of SPhP modes is that the dielectric permittivity $\varepsilon(\omega)$ near and within the Reststrahlen band, where $Re[\varepsilon(\omega)] \leq 0$, can be altered by introducing free carriers. This can be achieved either through electrostatic gating or optical pumping, thereby enabling the potential for *active tuning* of the SPhP modes (both traveling and localized resonator-types) using the LO-phonon free-carrier plasma coupling (LOPC) effect.[13, 16-21] The dielectric permittivity of polar dielectrics (e.g., 4H-SiC) can be expressed as a combination of a phononic term (in parentheses in Eq 1.) and a Drude term (summation in Eq. 1):[13]

$$\varepsilon(\omega) = \varepsilon_\infty \left(1 + \frac{\omega_{LO}^2 - \omega_{TO}^2}{\omega_{TO}^2 - \omega^2 - i\omega\gamma}\right) - \sum_{j=e,h} \frac{N_j e^2}{\varepsilon_0 m_j^* m_o} \frac{1}{\omega^2 + i\omega\Gamma} \tag{1}$$

where $\varepsilon_\infty$ is the high-frequency permittivity, $\omega_{LO}$ is the LO phonon frequency, $\omega_{TO}$ is the TO phonon frequency, $\omega$ is the incident frequency, and $\gamma$ is the damping constant. The free-carrier 'Drude' contribution (subscripts *e* and *h* denote electrons and holes), is defined by the $N$, $e$, $m^*$, $m_0$, $\varepsilon_0$, and $\Gamma$, which refer to the free-carrier density, electron charge, effective mass of the free carrier, the mass of an electron, the vacuum permittivity, and the plasmon damping constant, respectively.[12, 22] For all work discussed here, we take $N_e = N_h \equiv N$ because the photo-injected carriers far outnumber background carrier concentration that is ~$10^{14}$ cm$^{-3}$. Controlling the permittivity can therefore be achieved by altering $N$ via traditional doping, photo-injection or electro-modulation,[16] as shown in Fig. 1, which plots the real part of the dielectric function for 4H-SiC in the presence of excess free carriers using typical 4H-SiC optical constants.[13] Within the electrostatic limit, the resonant frequency of a polaritonic nanoparticle (plasmon or phonon) occurs at a definite negative value of the permittivity depending on particle shape.[22] For a nanosphere in air, the resonant condition is Re[ε(ω)] = -2. Consulting Fig. 1, where this condition is plotted as a dashed line, one sees that for a spherical particle, its associated local SPhP resonance will blue shift from approximately 10.75 μm (930 cm$^{-1}$) in the absence of free carriers to 10.3 μm (971 cm$^{-1}$) at carrier densities of approximately 1x10$^{19}$ cm$^{-3}$. Achieving such a shift with electro- or photo-modulation of free carriers would represent a dynamic tunability, which cannot be reasonably achieved with SPPs in most plasmonic materials other than graphene,[23-24] and could potentially facilitate a plethora of mid-IR to THz active nanophotonic applications.

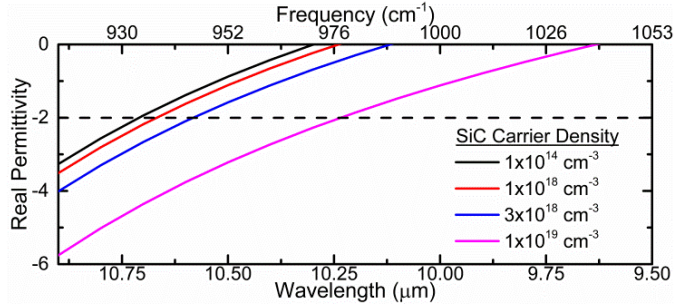

**Figure 1.** The tunable nature of SiC permittivity as a function of photo-injected carrier density. The dashed line represents the spherical resonance condition in air.

Here we demonstrate the dynamic tunability of the permittivity for thick, epitaxially grown 4H-SiC in the vicinity of the Reststrahlen band through the photo-injection of carriers. We use nanosecond pump-probe experiments to measure the free carrier induced reflection modulation arising from the LOPC effect. Large transient changes in reflectance are observed, increasing by nearly an order of magnitude at certain probe frequencies. This work differs in important ways from previous studies of the LOPC effect with ionized impurity doping.[19] First, in our work, the increased carrier density is transient in nature, which allows for fast carrier modulation required for actively tunable SPhP devices. Second, photo-injected carriers may be subject to reduced scattering, specifically with respect to the losses induced by the presence of ionized dopant atom incorporation prevalent with chemical doping. The results reported here exhibit rich spectral dynamics which provide an initial experimental demonstration of the photo-tunable nature of the permittivity in the vicinity of the Reststrahlen band. By fitting our experimental results, we provide theoretical predictions on the anticipated magnitude of the spectral shift and induced losses under the active modulation of a spherical resonator. These experiments hence provide an initial step towards realizing actively tunable SPhP-based photonics.

**Results and Discussion**

Measurements were carried out on an nominally undoped 160 μm thick 4H-SiC epitaxial layer (N ~ $10^{14}$ cm$^{-3}$) grown on a highly doped 4H-SiC substrate (N ~ 3 x$10^{18}$ cm$^{-3}$). The free carriers were injected into the epitaxial layer with a 355 nm pulsed laser, which has a penetration depth of approximately 50 μm,[25] thereby ensuring all measured signals result only from the epilayer and are not coming from the underlying substrate. We employed a variable pump-fluence $F$ ~ 1 to 34 mJ/cm$^2$, to provide a range of maximum absorbed photon densities $\Delta N_{max}$ from ~3x$10^{17}$ to 1x$10^{19}$ cm$^{-3}$ [$\Delta N_{max} = \alpha(1-r)F/h\nu$, where $h$ is Planck's constant and $\alpha$ = 210 cm$^{-1}$, $r$ = 0.15, and $\nu$ = 845 THz are the absorbance, reflectance, and pump frequency.[25]] The transient reflectance was probed with a continuous-wave, line-tunable CO$_2$ laser coupled with a fast mercury-cadmium-telluride detector that set the system temporal resolution to less than 100 ns. The pump and probe beams were both s-polarized and with angles of incidence of ~ 12° and 7° and beam diameters of 1.5 mm and 0.4 mm, respectively. The static reflectance spectrum was acquired with a FTIR spectrometer under identical angular and polarization conditions as the probe. The static and transient reflection spectra were fitted by using Eq. 1 in WVASE ellipsometry software from J.A. Woolam Inc. to extract estimates of the free-carrier densities.[26]

The static SiC reflection spectrum near the high-frequency (LO phonon) side of the Reststrahlen band is presented in Fig. 2 (black line) and demonstrates the high reflectivity that results from the screening field established by the polar optical phonons.[12, 22] Changes in the SiC reflectance spectrum due to the injection of free carriers were probed at the wavelengths denoted in Fig. 2 by the sets of dots that form coarse transient spectra. The red, green, and blue dots represent pump fluences of 1.0, 6.4, and 14.3 mJ/cm$^2$, corresponding to instantaneous absorbed photon-densities of $\Delta N_{max}$ ~ 0.3x$10^{18}$, 2.0x$10^{18}$, and 4.6x$10^{18}$ cm$^{-3}$, respectively. We note that immediately after excitation, the free carriers will begin to recombine through Auger and surface recombination-processes faster than our detector temporal resolution. Therefore, the reflectance value at each fluence is actually the value induced by a free-carrier density somewhat lower than the maximum instantaneous value. The experimental data points presented in Fig. 2 were extracted from the transient reflectance trace, $\Delta R(t)/R_o$ (see Fig. 3 for representative $\Delta R(t)/R_o$ traces), for each probe frequency at time $t$ ~100 ns (earliest time resolvable by our system) after the pump excitation. These data points were then scaled relative to the static-reflectance baseline.

The magnitudes of the transient reflectance signals exhibit a strong dependence on pump fluence and the sign of the change depends upon the probe frequency. As demonstrated in Fig. 2, for increasing pump fluence, the reflectance decreases for probe frequencies within the Reststrahlen band, while for frequencies along the band edge (~970 to 990 cm$^{-1}$) the reflectivity increases dramatically—by approximately an order of magnitude near 986 cm$^{-1}$. This "softening" and shifting of the Reststrahlen band edge to higher frequencies (i.e., the reduced slope near the high-frequency edge) results from the coupling between the polar LO phonons and the free-carrier-induced plasma (i.e., the LOPC effect).[13, 19]

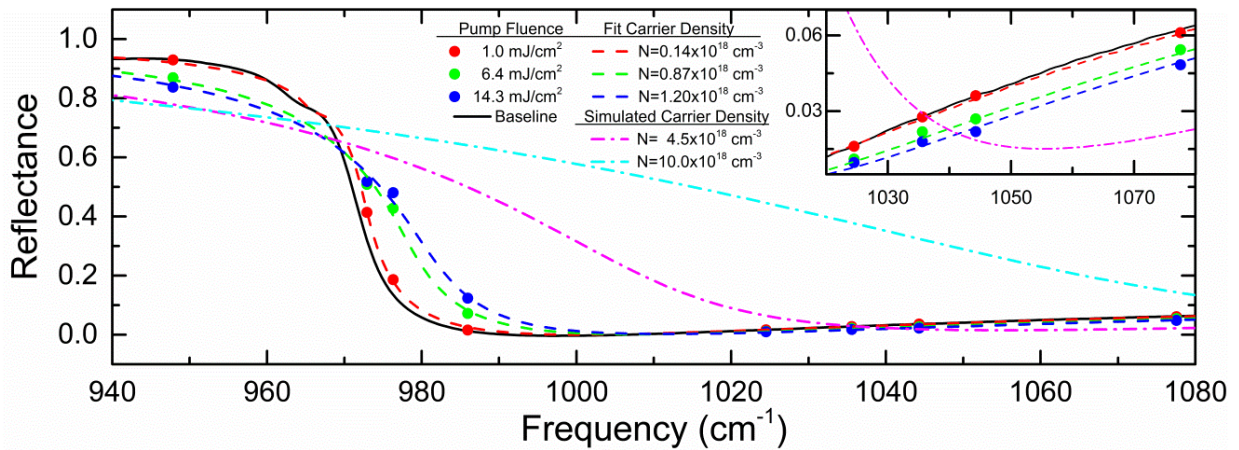

**Figure 2.** Reflectance spectra of SiC following photo-injection (355 nm) at pump fluences of $F \sim$ 1.0 mJ/cm$^{-2}$ (red dots) 6.4 mJ/cm$^{-2}$ (green dots) and 14.3 mJ/cm$^{-2}$ (blue dots) scaled relative to the baseline reflectance of the SiC sample (black line). Reflectance values were taken at the peak of the transient decay (~ 100 ns after excitation). The color coordinated dashed lines are reflectance-spectra fits for the respective carrier densities ($N$) of the experimental dots. The dash-dot lines represent calculated reflectance curves for 4.5x10$^{18}$ cm$^{-3}$ and 1x10$^{19}$ cm$^{-3}$ carrier densities. The inset shows the 1020 cm$^{-1}$ to 1080 cm$^{-1}$ region rescaled.

Overlaid on the experimental measurements in Fig. 2 are calculated fits (dashed lines) to the set of transient reflectance points (dots). For the fits, we employed the software WVASE, which computes reflectance with Fresnel equations incorporating the permittivity [Eq. (1)] in the appropriate tensor form for uniaxial 4H-SiC. A full list of the fitting values, as well as more details of the fitting procedure are discussed in the Supplemental Material. At $t \approx 100$ ns, the Drude term values determined from the fit were $N = 1.4$x$10^{17}$, $8.7$x$10^{17}$, and $1.2$x$10^{18}$ cm$^{-3}$ and $\Gamma = 2.4$x$10^{13}$, $7.7$x$10^{13}$, and $8.0$x$10^{13}$ s$^{-1}$ for the three fluences depicted by the red, green, and blue dashed lines in Fig. 1, respectively. As expected, these fitted densities fall below the maximum absorbed photon-densities of $\Delta N_{max} = 3.0$ x$10^{17}$, $2.0$x$10^{18}$, and $4.6$x$10^{18}$ cm$^{-3}$ (red, green, and blue dots) due to initial relaxation processes (tens of nanoseconds) that are not fully temporally resolved within our experimental setup.[25]

Taking a closer look at the dynamics, Fig. 3, we observe markedly different transient-reflectance behaviors within, on the edge of, and to the high-frequency side of the Reststrahlen band. Within the Reststrahlen band [probe frequency of 947.9 cm$^{-1}$; Fig. 3A], the reflectance trace shows a monotonic decrease in reflectance as the pump power increases and all traces show a single exponential decay (~ 700 ns) due to the subsequent recombination. We also determined the same ~700 ns exponential decay from free carrier absorption measurements probed at 7692 cm$^{-1}$ (1.3 μm), far removed from the Reststrahlen band, for every pump fluence (decays shown in the Supplementary Material). This exponential decay rate is consistent with measurements made previously on this sample by Ščajev et al.,[25] who attribute it to Shockley-Read-Hall (SRH) recombination due to deep traps together with surface recombination.[25] Observation of the rapid nonlinear Auger-decay expected for our higher injected carrier densities[25] is precluded by our detector time resolution. However, the presence of this decay mechanism can be inferred from the fluence dependence of the normalized maximum change in reflectance at 947.9 cm$^{-1}$ [filled black circles in the inset to Fig. 3A], which shows a linear trend for low $\Delta N_{max}$, followed by a sub-linear 'roll-off' at higher pump densities. We find a very similar fluence-dependence for the maximum change in transmittance in independent free carrier absorption measurements performed at 7692 cm$^{-1}$, (open red circles in the inset).[25] The nonlinear behavior is a direct result of interband Auger recombination (for which the recombination rate scales as $N^2$),[25] an established recombination pathway in 4H-SiC for carrier densities in excess of ~ $10^{18}$ cm$^{-3}$, similar to those used here.[25] Based on the strong similarity between the two sublinear sets of data in the inset of Fig. 3A measured at short time, and the correctly inferred exponential decay-time of 700 ns at longer times, we conclude that carrier dynamics can be probed directly using changes in reflectance within the highly reflective region of the Reststrahlen band. Further rationale is given in Supplemental Material.

Unlike the transient decay rates measured within the Reststrahlen band at 947.9 cm$^{-1}$, which are independent of carrier density, those measured on the steep high-frequency edge of the band at 976.6 cm$^{-1}$ [Fig. 3B] exhibit a functional form that is non-exponential and carrier-density dependent at early times. In addition, the reflectance here nearly tripled at the highest pump fluence. Similarly, at the ~986 cm$^{-1}$ probe (near the baseline reflectance minimum in the spectra in Fig. 2,) the reflectance jumped nearly an order of magnitude at the highest pump fluence. Nagai et al.[27] observed similar dynamical changes in reflectance for GaN (> 400%) using ultrafast techniques with finer time resolution. The complex decays and extreme reflectance changes are due to the high-frequency edge of the Reststrahlen band sweeping to higher energy, a consequence of the hybridized LOPC mode mediated by the photo-injected carrier density *N*. This hybridized mode is comprised of upper ($\omega_+$) and lower ($\omega_-$) branches. As *N* is increased, $\omega_+$ (near the LO phonon frequency for low *N*) is driven to higher frequencies as it approaches the avoided crossing (defined by the plasma frequency), forcing significant changes in reflectance.[28] Harima et al. demonstrated this effect *statically* using Raman scattering of SiC epitaxial layers with different ionized dopant levels,[13, 19] while others have used modified graphene overlayed on SiC to *statically* tune the permittivity.[29-30] Our results indicate that the decay dynamics are non-exponential at longer times (after nonlinear Auger becomes ineffective) because as the carriers recombine through the SRH and surface recombination mechanisms, the minimum in reflectance moves back from the blue-shifted excited state, approaching the uncoupled LO phonon position. This dynamic contribution from the carrier dependent LOPC and various simultaneous recombination mechanisms provide for the complicated nonlinear response of the probe, making deconvolution of each phenomenon difficult. Nevertheless, the dynamic photo-tuning of reflectance caused by the carrier-modulated permittivity is clearly demonstrated.

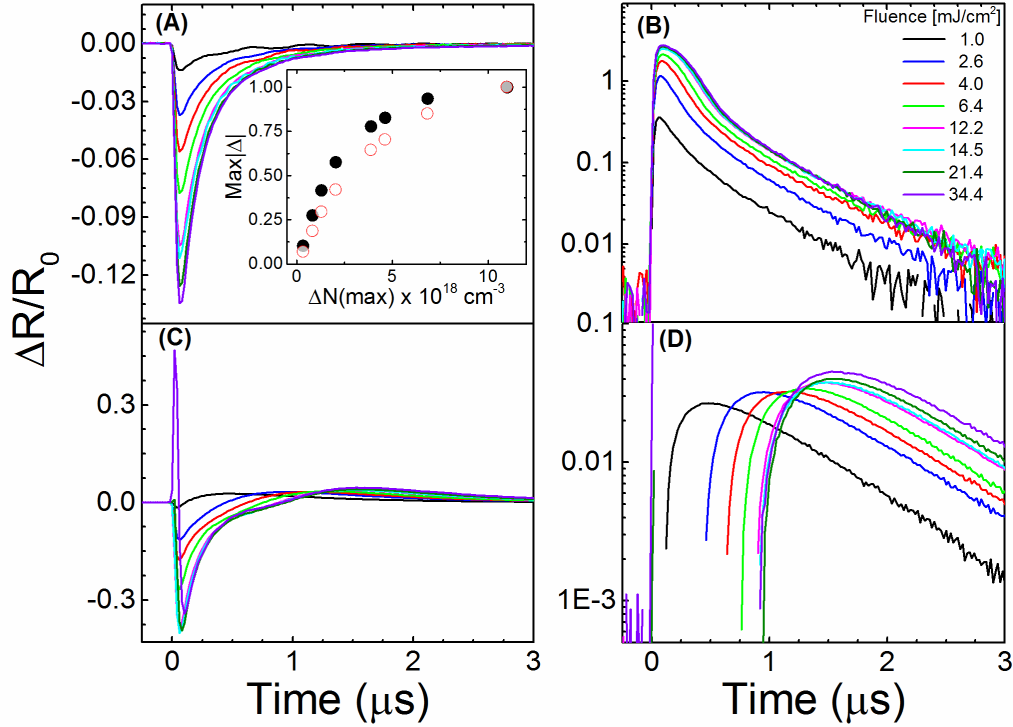

**Figure 3**. Transient reflectance dynamics of SiC for various probe wavelengths (A: 947.9 cm$^{-1}$; B: 976.6 cm$^{-1}$; C: 1043.8 cm$^{-1}$) near the blue edge of the Reststrahlen band. The pump fluence was varied from $F \sim 1.0$ to 34.4 mJ/cm$^2$, corresponding to photon densities of $\Delta N_{max} \sim 0.33$ to $11.0 \times 10^{18}$ cm$^{-3}$, for each probe wavelength. The inset of panel A shows the normalized peak change in reflectance at 947.9 cm$^{-1}$ (filled black circles) and free carrier absorption at 7692 cm$^{-1}$ (open red circles) as a function of photo-injection density. Panel D shows a semi-log plot of the transient reflectance at 1043.8 cm$^{-1}$ (from panel C).

More complicated transient behavior is observed in the higher energy regions blue of the Reststrahlen band edge. In particular, at a probe frequency of 1043.8 cm$^{-1}$ [Fig. 3C] at the highest photon density (i.e., $\Delta N_{max} > 1 \times 10^{19}$ cm$^{-3}$), $\Delta R(t)/R_o$ is initially positive, demonstrating a transient increase in reflectance. However, the sign of the transient reflectance rapidly becomes negative as the carrier density decreases. To account for this behavior, we calculated the reflectance spectra at the high carrier densities of $4.5 \times 10^{18}$ and $1.0 \times 10^{19}$ cm$^{-3}$ (magenta and cyan dash-dotted lines in Fig. 2, respectively). In the 1030 to 1040 cm$^{-1}$ frequency range (see inset of Fig. 2), the calculations predict at the higher concentration that the reflectance should sharply increase, but as the carrier density decays below roughly $4 \times 10^{18}$ cm$^{-3}$ the reflectance should transition to a reflectance decrease, in qualitative agreement with our experiments. We note, however, that with further free carrier recombination, our experiments exhibit $\Delta R(t)/R_o$ once again becoming positive for ~ t>200 ns [Fig. 3C], a behavior not accounted for in our permittivity model. Eventually, the transient reflectance decays exponentially back to the baseline with a time constant of ~700 ns [see Fig. 3D] matching the free carrier absorption dynamics discussed above.

Our analysis of the transient reflectance measurements provides insight into free carrier induced changes in the dielectric permittivity $\varepsilon(\omega)$. In our experiments, the positive spike feature in Fig. 3C is consistent with our calculations that free-carrier densities ($N$) of order $10^{19}$ cm$^{-3}$ were photo-injected, even if they could not be directly probed temporally. Based on the modeled permittivity traces presented in Fig. 1, it is predicted that a spherical SiC resonator (dashed line in Fig. 1) would exhibit a spectral shift ($\delta\omega$) in excess of 40 cm$^{-1}$ due to such a high free carrier injection level. Such broad tuning of a resonance by rapid injection of carriers would be useful for a range of architectures, perhaps leading to actively tunable mid-infrared nanoscale optical antennas, waveguides, or epsilon-near-zero (ENZ) behavior.[16]

While large spectral tuning is expected at high carrier densities, there is an important caveat to consider with regard to linewidth broadening of spectral resonances. As the carrier density increases and the Drude terms in Eq. 1 begin to dominate the dielectric function, one may expect broadening of the resonance due to the shorter scattering times of electrons compared to phonons. To determine the functionality of SPhP resonance tuning using carrier injection, we consider the simple model system of a spherical polar dielectric resonator with a radius *a* that is much smaller than the incident free-space wavelength ( $\lambda$ ). The ratio of the sphere's absorption cross-section to its geometric cross-section is given by the following:[22]

$$Q_{abs} = \frac{24\pi a}{\lambda} \frac{\varepsilon_m^{3/2} \varepsilon''}{(\varepsilon' + 2\varepsilon_m)^2 + \varepsilon''^2} . \qquad (2)$$

Here, $Q_{abs}$ is the absorption efficiency, not to be confused with the quality factor Q, with $\varepsilon_m$, $\varepsilon'$, and $\varepsilon''$ representing the dielectric constant of the surrounding medium, and the real and imaginary part of the complex dielectric function of the polar dielectric, respectively. In order to clearly illustrate our point, we make a simplifying approximation of an isotropic permittivity given by $\varepsilon = \varepsilon' + i\varepsilon''$. While 4H-SiC is technically a uniaxial crystal, previous studies have used this approximation to describe localized SPhP modes with excellent agreement.[2, 5-6] Using the real and imaginary parts of the dielectric permittivity derived from Fig. 2 for each carrier density, we show the potential active tuning of a SiC spherical nanoresonator in air by plotting $Q_{abs}$ in Fig. 4A. As shown by Fig. 4, linewidth broadening accompanies the shifting of the peak because of the corresponding fast scattering rates of the photo-injecting carriers. This is displayed more quantitatively in Fig. 4B, which demonstrates that as the carrier density increases the resonance exhibits a blue-shift in the peak position along with increasing FWHM values. The additional broadening resembles results from previous Raman studies of doped SiC, which show apparent shifts and broadening of the LO phonon band with increasing carrier concentration.[19-20] Based on the estimated permittivities derived from Fig. 2, we would expect a spectral shift, $\delta\omega$, of about 5 cm$^{-1}$ at a injected carrier density of 1.2 x 10$^{18}$ cm$^{-3}$ with the resonance linewidth, $\Delta\omega = 8$ cm$^{-1}$, corresponding to a modulation depth of $\delta\omega / \Delta\omega = 0.6$ and a quality factor of $\omega/\Delta\omega = 120$. Therefore, even given the additional broadening, the quality factor is still significantly higher than any reported plasmonic systems (a survey of such Q factors in both SPP and SPhP materials may be found in Ref. [11]), and the modulation depth is comparable to the highest tuning ranges reported for plasmonic systems.[31-32] Additionally, a major benefit of the tuned SPhPs is that their linewidths are much narrower than their plasmonic counterparts, so the former could potentially be applied to modulated surface enhanced IR (SEIRA) techniques.

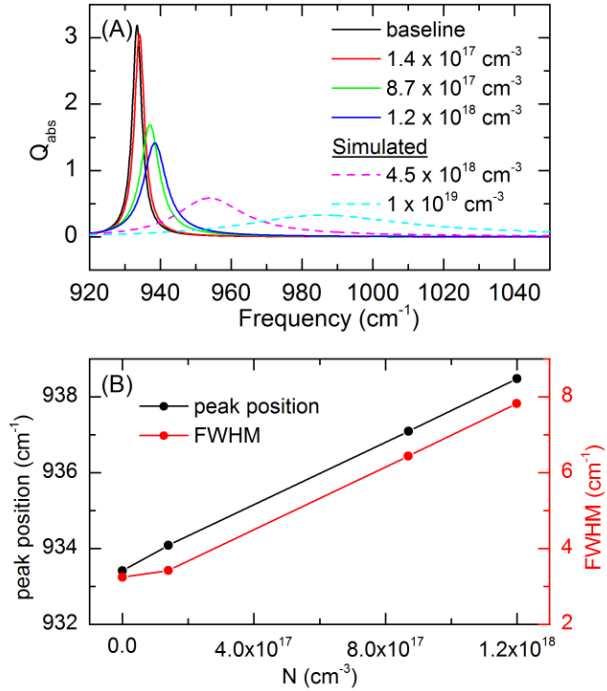

**Figure 4**. Theoretical spherical surface phonon polarion resonance tuning of 50 nm radius SiC nanoresonator. Solid lines in panel A are calculated from the permittivity values derived from Fig. 2. The lines are labeled based on the estimated carrier density. The dash lines in panel A represent simulated tuning for carrier densities of $4.5 \times 10^{18}$ and $1 \times 10^{19}$ cm$^{-3}$. Panel B plots the resonant peak positions and FWHM values of the experimental data from this study (solid lines in panel A).

As mentioned above, the temporally unresolved positive feature in Fig. 3C suggests we are approaching carrier densities of $\sim 10^{19}$ cm$^{-3}$. Fig. 4A includes calculated curves for carrier densities of $4.5 \times 10^{18}$ and $1.0 \times 10^{19}$ cm$^{-3}$ (dashed lines) to show the potential tunability of the resonances at higher carrier densities. For these two higher carrier densities, we let the free-carrier-induced damping be fixed at the maximum value derived from our fits, i.e., $\Gamma = 8.0 \times 10^{13}$ s$^{-1}$. At the highest carrier density of $1 \times 10^{19}$ cm$^{-3}$, the resonance shifts $> 40$ cm$^{-1}$. Note, however, that even though the Drude damping parameter was kept constant for the carrier densities of $1.2 \times 10^{18}$ cm$^{-3}$, $4.5 \times$ $10^{18}$ cm$^{-3}$, and 1 x $10^{19}$ cm$^{-3}$ (which correspond to plasma frequencies of 506 cm$^{-1}$, 980 cm$^{-1}$, and 1461 cm$^{-1}$), the resonance linewidth still shows significant broadening for the highest carrier densities. This large increase in linewidth is relatively insensitive to the value of the damping parameter. For example, doubling the Drude damping parameter from 8.0 x $10^{13}$ s$^{-1}$ to 16 x $10^{13}$ s$^{-1}$ only increases the linewidth from ~ 55 cm$^{-1}$ to ~70 cm$^{-1}$. This large increase in resonance linewidth at higher carrier densities occurs because the Drude term begins to dominate the SiC dielectric function as the plasma frequency approaches $\omega_{LO}$. At a carrier density of 1 x $10^{19}$ cm$^{-3}$ (4.5 x $10^{18}$ cm$^{-3}$), we calculate a modulation depth of ~ 1 (0.9) and a quality factor ~ 18 (42). Therefore, even with significant broadening due to increased carrier-carrier scattering from the photo-injection, there is potential to actively tune nanoresonators over large spectral regions while maintaining comparable figures of merit to traditional plasmonic systems.

While we have demonstrated the carrier-induced transient tuning of the dielectric function of 4H-SiC through optical pumping, this effect can also be potentially realized through carrier injection within electronic devices, for instance gated transistor architectures or bipolar electronic devices (e.g. pin diodes). When coupled with the controllable carrier recombination times ranging from the picoseconds to tens of microseconds[33] in direct and indirect band gap semiconductors, respectively, the active tunability provided by carrier injection demonstrated here presents a clear path towards modulated and active photonic devices in the mid-infrared to THz spectral ranges.

**Concluding Remarks**

In conclusion, we have employed transient infrared pump-probe spectroscopy with variable levels of photo-carrier injection to explore the effects of free carriers in the Reststrahlen band region of SiC. The pump-probe experiments revealed complex transient behavior that provides evidence for the LOPC-mediated carrier-tunable permittivity of 4H-SiC. By fitting the transient reflectance spectra with a model of permittivity that included phononic and free-carrier terms, we are able to extract estimates of the free-carrier densities and carrier-induced damping under various optical pumping levels. The inferred carrier-induced changes in the dielectric permittivity enable the prediction of the potential for active tunability of SPhP architectures. For example, at our highest pump fluence (34 mJ/cm$^2$) the reflectance behavior implies a carrier density of ~$10^{19}$ cm$^{-3}$, which predicts a ~ 40 cm$^{-1}$ shift for the local SPhP resonance in a SiC nanosphere. This suggests prospective photo-modulation in these polar dielectrics over an order of magnitude greater than the modulation depth of other active nanophotonic approaches. Hence, the results provided here indicate the potential for achieving dynamic tunability of localized SPhP modes in sub-diffractional optical antennas,[2, 5-6, 8] waveguides,[12, 16] and ENZ based effects. Such functionality would provide a novel approach towards realizing active nanophotonics in the mid-IR to THz frequency regions at modulation frequencies from the kHz to GHz.

**Supporting Information.** Additional information on experimental details, calculations and corresponding assumptions, and free carrier absorption data to support results in the main text. This material is available free of charge via the Internet.

**AUTHOR INFORMATION**

[†]Authors contributed equally to this work

**Corresponding Authors:** Joshua D. Caldwell, *email: joshua.caldwell@nrl.navy.mil, and Jeffrey C. Owrutsky, *email: jeff.owrutsky@nrl.navy.mil

## ACKNOWLEDGEMENTS

The authors wish to acknowledge support from the U.S. Office of Naval Research through the Naval Research Laboratory Nanoscience Institute. B.T.S, R.C., A.D.D., and A.J.G. thank the National Research Council for administering their postdoctoral fellowships at NRL. The authors would like to thank Igor Vurgaftman (NRL), Chris Kendziora (NRL), Orest Glembocki (NRL), and Jon Schuller (Univ. California – Santa Barbara) for helpful discussions.
## REFERENCES

1. Maier, S. A., *Plasmonics: fundamentals and applications: fundamentals and applications*. Springer Science & Business Media: 2007.
2. Caldwell, J. D.; Glembocki, O. J.; Francescato, Y.; Sharac, N.; Giannini, V.; Bezares, F. J.; Long, J. P.; Owrutsky, J. C.; Vurgaftman, I.; Tischler, J. G., Low-loss, extreme subdiffraction photon confinement via silicon carbide localized surface phonon polariton resonators. *Nano letters* **2013,** *13* (8), 3690-3697.
3. Hillenbrand, R.; Taubner, T.; Keilmann, F., Phonon-enhanced light-matter interaction at the nanometre scale. *Nature* **2002,** *418* (6894), 159-162.
4. Greffet, J.-J.; Carminati, R.; Joulain, K.; Mulet, J.-P.; Mainguy, S.; Chen, Y., Coherent emission of light by thermal sources. *Nature* **2002,** *416* (6876), 61-64.
5. Chen, Y.; Francescato, Y.; Caldwell, J. D.; Giannini, V.; Maß, T. W.; Glembocki, O. J.; Bezares, F. J.; Taubner, T.; Kasica, R.; Hong, M., Spectral tuning of localized surface phonon polariton resonators for low-loss mid-ir applications. *ACS Photonics* **2014,** *1* (8), 718-724.
6. Wang, T.; Li, P.; Hauer, B.; Chigrin, D. N.; Taubner, T., Optical properties of single infrared resonant circular microcavities for surface phonon polaritons. *Nano letters* **2013,** *13* (11), 5051-5055.
7. Dai, S.; Fei, Z.; Ma, Q.; Rodin, A.; Wagner, M.; McLeod, A.; Liu, M.; Gannett, W.; Regan, W.; Watanabe, K., Tunable phonon polaritons in atomically thin van der Waals crystals of boron nitride. *Science* **2014,** *343* (6175), 1125-1129.
8. Caldwell, J. D.; Kretinin, A. V.; Chen, Y.; Giannini, V.; Fogler, M. M.; Francescato, Y.; Ellis, C. T.; Tischler, J. G.; Woods, C. R.; Giles, A. J., Sub-diffractional volume-confined polaritons in the natural hyperbolic material hexagonal boron nitride. *Nature communications* **2014,** *5*.
9. Xu, X. G.; Ghamsari, B. G.; Jiang, J.-H.; Gilburd, L.; Andreev, G. O.; Zhi, C.; Bando, Y.; Golberg, D.; Berini, P.; Walker, G. C., One-dimensional surface phonon polaritons in boron nitride nanotubes. *Nature communications* **2014,** *5*.
10. Holmstrom, S. A.; Stievater, T. H.; Pruessner, M. W.; Park, D.; Rabinovich, W. S.; Khurgin, J. B.; Richardson, C. J. K.; Kanakaraju, S.; Calhoun, L. C.; Ghodssi, R., Guided-mode phonon-polaritons in suspended waveguides. *Physical Review B* **2012,** *86* (16), 165120.
11. Caldwell, J. D.; Lindsey, L.; Giannini, V.; Vurgaftman, I.; Reinecke, T.; Maier, S. A.; Glembocki, O. J., Low-Loss, Infrared and Terahertz Nanophotonics with Surface Phonon Polaritons. *Nanophotonics* **2015,** *4* (1), 44-68.
12. Caldwell, J. D.; Lindsay, L.; Giannini, V.; Vurgaftman, I.; Reinecke, T. L.; Maier, S. A.; Glembocki, O. J., Low-loss, infrared and terahertz nanophotonics using surface phonon polaritons. *Nanophotonics* **2014,** *3*, 2192-8606.
17

**For table of contents use only**

**Table of contents figure**

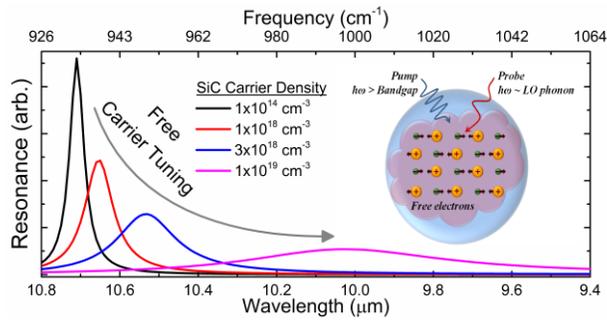

# Supplementary material: Photoinduced tunability of the Reststrahlen band in 4H-SiC

Bryan T. Spann,[1,2,†] Ryan Compton,[1,2,†] Daniel Ratchford,[1] James P. Long,[1] Adam D. Dunkelberger,[1,2] Paul B. Klein,[3] Alexander J. Giles,[1,2] Joshua D. Caldwell,[1,*] and Jeffrey C. Owrutsky[1,*]

[1]U.S. Naval Research Laboratory, Washington, D.C., 20375, USA

[2]National Research Council Postdoctoral Research Associate

[3]Sotera Defense Solutions Inc., Herndon, Virginia, 20171, USA

[†]Authors contributed equally to this work

[*]Corresponding authors: joshua.caldwell@nrl.navy.mil, jeff.owrutsky@nrl.navy.mil

**Supplemental Experimental Details**

Measurements were carried out on a 160 μm thick 4H-SiC epitaxial layer with a background n-type carrier density of ~$10^{14}$ cm$^{-3}$ grown on a standard highly doped, n$^+$ 4H-SiC substrate ($N_d$~$3\times10^{18}$ cm$^{-3}$) that was 8° miscut off the {0001} plane toward <1120>. The free-carriers were excited with the third harmonic (λ = 355 nm) of a pulsed Nd:YAG laser (4 ns, 20 Hz; New Wave Polaris II). The reflectance was probed with a continuous-wave line-tunable CO$_2$ laser (California Laser). The pump and probe beams were both s-polarized and with angles of incidence of ~ 12º and 7º, with beam diameters ~1.5 mm and ~0.4 mm, respectively [see Fig. S1 for the pump-probe schematic (A) and sample orientation (B)]. The reflected probe-power $R(t)$ was monitored with a fast mercury-cadmium-telluride detector that set the system temporal resolution of less than 100 ns. For each probe wavelength, the initial SiC reflected power, $R_o$, was measured before application of the pump laser. The relative transient-reflectance trace, $\Delta R(t)/R_o$, was calculated as $\Delta R(t) = R(t) - R_o$ normalized by $R_o$. The static-reflectance baseline spectrum $R_b$ was acquired with an FTIR spectrometer under identical angular and polarization conditions as the probe. We used a gold sample as a reference for the FTIR spectrum, however, because SiC has a higher reflectance than gold in the spectral region under investigation, we normalized the maximum reflectance to ~99 %. See the red dashed line of Figure S2 for the reflectance over the entire Reststrahlen band range along with the calculated real part of the permittivity (solid black line of Fig. S2).

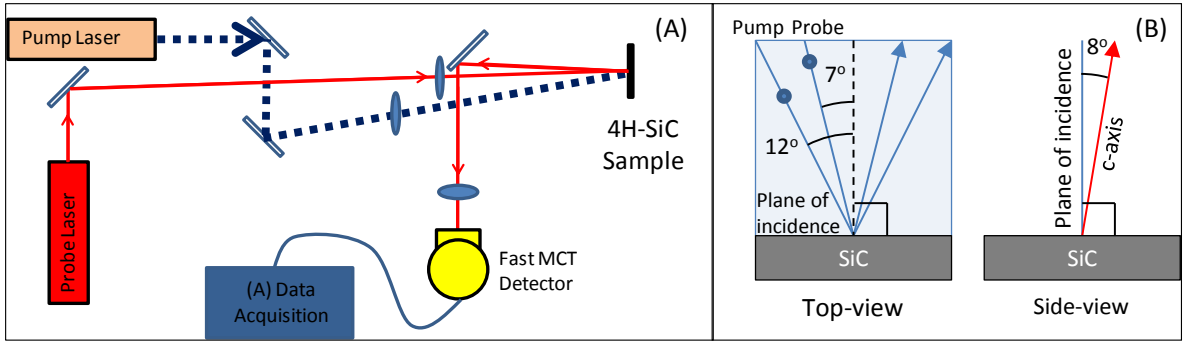

**Figure S1.** (A) Pump-probe experimental schematic. (B) Detailed orientation of the sample.

Additionally, we note that the instantaneous, absorbed photon densities ($\Delta N_{max}$) represent the theoretical maximum values for the electron and hole densities $N_e$ and $N_h$ generated by the pump pulse, but that these carriers will begin to recombine through bulk-Auger and surface recombination-processes during our detector temporal resolution. The experimental data points presented in Fig. 2 for each pump fluence and each probe frequency were extracted from the transient-reflectance trace $\Delta R(t)/R_o$ at time $t \sim 100$ ns after the pump excitation (the time nearest to the initial injection while still being outside the instrument response time). To convert these relative changes to actual reflectance, these data points were then scaled relative to the measured static-reflection baseline, i.e., $R(\Delta N_{max}) = R_b[\Delta R(t = 100\ ns)/R_0 + 1]$.

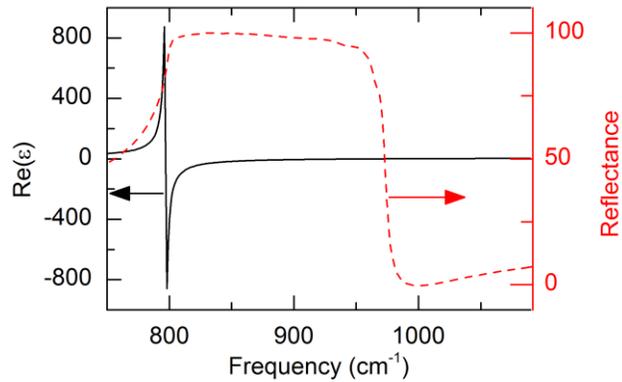

**Figure S2.** Real part of the permittivity (solid black line) and reflectance (red dashed line) spanning full Reststrahlen band spectral region.

**Supplementary Simulation Details and Experimental Results**

For the fits, we employed the software WVASE, which computes reflectance with Fresnel equations incorporating the permittivity of Eq. (1) in tensor form. For 4H-SiC (a uniaxial crystal), the permittivity tensor is given by $\varepsilon_{xx} = \varepsilon_{yy} = \varepsilon^t$ along the transverse crystallographic x and y axes and $\varepsilon_{zz} = \varepsilon^z$, along the optical z axis. We also applied the appropriate rotational transformation matrix to the permittivity tensor to account for the 8° offset of the c-axis from the surface normal. For simplicity, we assumed a spatially uniform carrier-distribution for $N_e$ and $N_h$ in Eq. 1, rather than the expected distribution that peaks 10 to 20 μm beneath the surface due to surface recombination, as reported in a study of similar samples.[1] Test computations in WVASE found that the reflectance spectrum for such non-uniform carrier-profiles closely matched the spectrum from a uniform distribution with a carrier density equal to the surface concentration. We also assumed that the injected carrier density is much larger than the density of carrier traps (measured at $2.7 \times 10^{13} cm^{-3}$)[2], so that the electron and hole carrier densities were approximately the same ($N = N_e \cong N_h$). The static baseline reflectance was initially fit using the above permittivity model without the Drude terms (red dashed line in Fig. S3). The intrinsic carrier density of the epitaxial layer is $<10^{15}$ cm$^{-3}$, and at such low concentrations the Drude term in the permittivity has a negligible contribution.[3-4] Since the transverse optical-phonon frequencies were outside the fitting region window, they were set to be $\omega_{TO}^z = 782 cm^{-1}$ and $\omega_{TO}^t = 797 cm^{-1}$.[5] All other parameters in the Lorentzian term were allowed to be free, and the resultant fit overlays the data well. The fit parameters for this scenario were found to be: $\varepsilon_\infty^t = 6.61$, $\varepsilon_\infty^z = 7.32$, $\omega_{LO}^t = 970.91 cm^{-1}$, $\omega_{LO}^z = 964.48 cm^{-1}$, $\gamma^t = 3.24 cm^{-1}$, $\gamma^z = 3.54 cm^{-1}$. The resultant fit is displayed in Fig. S3.

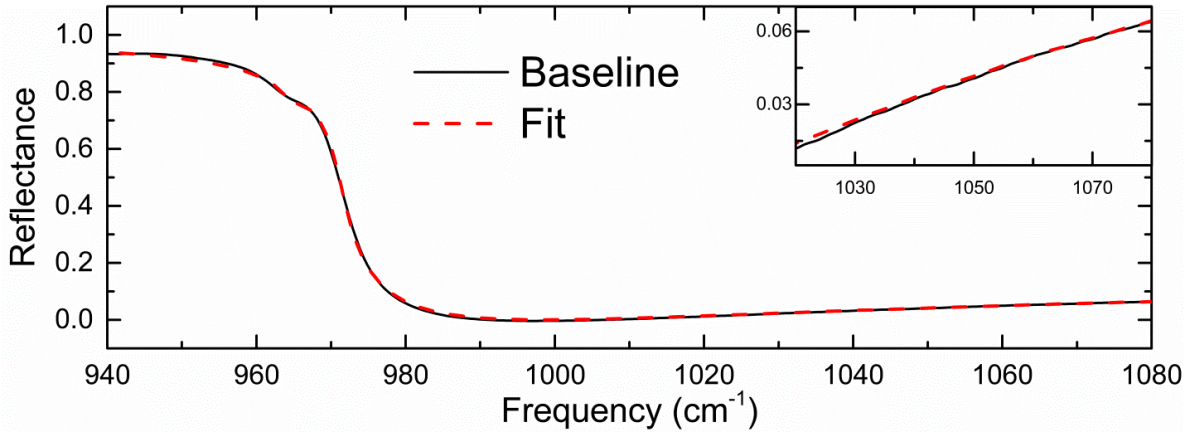

**Figure S3.** Baseline reflectance with respective fit from the WVASE software assuming no Drude term contribution.

These baseline fit values were subsequently used as fixed parameters in the Lorentzian oscillator terms of Eq. (1) to fit the transient data points. The transverse (t) component of the effective mass of the electrons (holes) was taken be 0.42[6] (0.66)[7], and the z component was taken be 0.29[6] (1.75)[7]. Therefore, the only free parameters used in fitting the transient spectra were $N$ and $\Gamma$. At $t \approx 100$nsec, the values determined from the fit were $N = 1.4 \times 10^{17}$, $8.7 \times 10^{17}$, and $1.2 \times 10^{18}$ cm$^{-3}$ and $\Gamma = 2.4 \times 10^{13}$, $7.7 \times 10^{13}$, and $8.0 \times 10^{13}$ s$^{-1}$ for the three fluences depicted by the red, green, and blue dashed lines in Fig. 2, respectively.

**Use of Transient Reflectance to Extract Carrier Dynamics and Supplementary Experimental Results**

In the main article, it was experimentally determined that the relative transient-reflectance, $\Delta R(t)/R_0$, provided a proportionate measure of the carrier density $N(t)$ to quite high densities (at least $10^{19}$ cm$^{-3}$), provided the measurement was made within the Reststrahlen band where the baseline reflectance approaches unity [see inset to Fig. 3(A)]. This can be explained by using the familiar Fresnel relation for the reflectance from a planar interface assuming normal incidence with a uniform medium under lossless conditions: $R = (n-1)^2/(n+1)^2 = (\sqrt{\varepsilon} - 1)^2/(\sqrt{\varepsilon} + 1)^2$, where $n$ is the refractive index, $\varepsilon$ is the optical permittivity, and we've used the relation $n = \sqrt{\varepsilon}$. Expanding to first order in $\varepsilon$, $\delta R/R = 2\delta\varepsilon/\sqrt{\varepsilon}(\varepsilon - 1)$. Therefore, assuming that $|\varepsilon| \gg 1$ (as it is within the highly reflective part of the Reststrahlen band), $\delta R/R$ is robustly linear with $\delta \varepsilon$ and hence to $\delta N$ (according to Eq. 1).

Transient free-carrier absorption decays measured in *transmission* with a probe frequency of 7692 cm$^{-1}$ (1.3 μm), much bluer than the Reststrahlen band, were used to confirm the 700 ns decay rate measured by transient *reflectance* near the Reststrahlen band. The transient transmission signal is shown below in Figure S4 for various fluences, and displays the same ~700 ns decay time measured in reflectance. Note that the change in transmission is shown as an absolute value in order to plot in log-scale, the change in transmission is actually negative because the free-carriers absorb the probe photons causing reduced transmission.

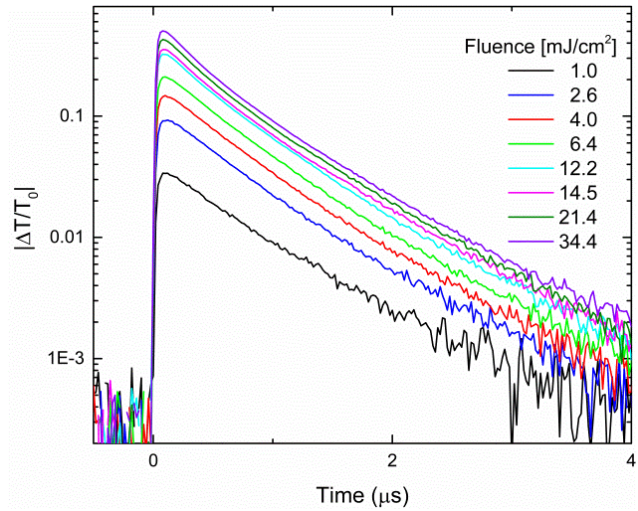

**Figure S4.** Transient free-carrier absorption measurements at a probe energy of 7692 cm$^{-1}$. The pump fluence was varied from $F$ ~ 1.0 to 34.4 mJ/cm$^2$, corresponding to photon densities of $\Delta N_{max}$ ~ 0.33 to 11.0 x 10$^{18}$ cm$^{-3}$, for each probe wavelength.

**Calculation of SiC nanosphere resonances**

Below is a table containing the permittivity values used to calculate $Q_{abs}$ in Fig 4.

| Carrier density: $N[cm^{-3}]$ | $\Gamma[s^{-1}]$ | $\varepsilon_\infty$ | $\omega_{TO}[cm^{-1}]$ | $\omega_{LO}[cm^{-1}]$ | $\gamma[cm^{-1}]$ | $m_e^*/m_h^*$ |
|---|---|---|---|---|---|---|
| Baseline, N = 0 | - | 6.61 | 797 | 970.91 | 3.24 | - |
| 1.4x10$^{17}$ | 2.4x10$^{13}$ | 6.61 | 797 | 970.91 | 3.24 | 0.42/1.2 |
| 8.7x10$^{17}$ | 7.7x10$^{13}$ | 6.61 | 797 | 970.91 | 3.24 | 0.42/1.2 |
| 1.2x10$^{18}$ | 8.0x10$^{13}$ | 6.61 | 797 | 970.91 | 3.24 | 0.42/1.2 |
| 4.5x10$^{18}$ | 8.0x10$^{13}$ | 6.61 | 797 | 970.91 | 3.24 | 0.42/1.2 |
| 1x10$^{19}$ | 8.0x10$^{13}$ | 6.61 | 797 | 970.91 | 3.24 | 0.42/1.2 |